\documentclass[reprint,superscriptaddress,showpacs,amsmath,amssymb,prb]{revtex4-1}
\usepackage{graphicx}% Include figure files
\usepackage{wrapfig}
\usepackage{epsfig}
\graphicspath{ {./images/} }
\usepackage{dcolumn}% Align table columns on decimal point
\usepackage{bm}% bold math
\usepackage{amsmath}
\usepackage{gensymb}
\usepackage{color}
\usepackage{epstopdf}
\usepackage{comment}
\usepackage{cuted}
\usepackage{mathtools}
\usepackage{lipsum}
\usepackage{upgreek}
\usepackage{natbib}
\usepackage{booktabs}
\usepackage{hyperref}% add hypertext capabilities
\usepackage[mathlines]{lineno}% Enable numbering of text and display math
\begin{document}

%\preprint{APS/123-QED}
\preprint{\today}

\title{Antiferromagnetic spin canting and Magnetoelectric Multipoles in h-YMnO$_{3}$}

\author{M. Ramakrishnan}
 \email{mahesh.ramakrishnan@hotmail.com}
 \affiliation{Swiss Light Source, Paul Scherrer Institut, 5232 Villigen PSI, Switzerland}
 \affiliation{MAX IV Laboratory, Lund University, PO Box 118, SE-22100 Lund, Sweden}
\author{Y. Joly}
 \affiliation{Universit\'{e} Grenoble Alpes, Institut N\'{e}el, F-38042 Grenoble, France}
 \author{Q. N. Meier}
 \affiliation{Department of Materials, ETH Zurich, 8093 Zurich, Switzerland}
\author{M. Fechner}
 \affiliation{Department of Materials, ETH Zurich, 8093 Zurich, Switzerland}
 \altaffiliation{Present address: DESY..., Germany}
\author{M. Porer}
 \affiliation{Swiss Light Source, Paul Scherrer Institut, 5232 Villigen PSI, Switzerland}
\author{S. Parchenko}
 \affiliation{Swiss Light Source, Paul Scherrer Institut, 5232 Villigen PSI, Switzerland}
\author{Y. W. Windsor}
 \altaffiliation{Present address: Department of Physical Chemistry, Fritz-Haber-Institut of the Max Planck Society, Faradayweg 4-6, Berlin 14915, Germany}
 \affiliation{Swiss Light Source, Paul Scherrer Institut, 5232 Villigen PSI, Switzerland}
\author{E. M. Bothschafter}
 \affiliation{Swiss Light Source, Paul Scherrer Institut, 5232 Villigen PSI, Switzerland}
\author{F. Lichtenberg}
 \affiliation{Department of Materials, ETH Zurich, 8093 Zurich, Switzerland}
\author{U. Staub}
 \email{urs.staub@psi.ch}
 \affiliation{Swiss Light Source, Paul Scherrer Institut, 5232 Villigen PSI, Switzerland}

%Ground-state atomic multipoles which simultaneously break parity and time-reversal symmetries have been discussed in condensed matter systems such as multiferroics and high-T$_C$ superconductors. We examine the possibility of the occurrence of these magnetoelectric multipoles in the archetypal multiferroic hexagonal YMnO$_3$ (h-YMO) using resonant x-ray diffraction. We observe an antiferromagnetic spin-canting in h-YMO, which has not been observed by diffraction thus far. We show that the peculiar evolution of the diffraction spectral profile at the Mn $L_{2,3}$ edges with temperature can be described by the interference of scattered signals from a magnetic dipole and a polar toroidal octupole.  

\begin{abstract}
Hexagonal YMnO$_{3}$ is a prototype antiferromagnet which exhibits multiferroic behavior with the ferroelectric and magnetic transitions occurring at different temperatures. We observe an out-of-plane canting of the Mn$^{3+}$ magnetic moments using resonant X-ray diffraction (RXD) in a single crystal of this material. These canted moments result in the symmetry-forbidden (0,0,1) magnetic Bragg reflection, which is observed at the Mn $L_{2,3}$ absorption edges. We also observe an unexpected difference in the RXD spectral shapes at different temperatures. Using \textit{ab initio} calculations, we explore the possibility that this behavior arises due to the interference between scattering from the canted magnetic moments and parity-odd atomic multipoles on the Mn$^{3+}$ ions.
\end{abstract}

% \pacs{71.15.Mb, 75.25.-j, 75.85.+t, 78.70.Ck}

\maketitle

%%%%%%%%%%%%%%%%%%%%%%%%%%%%%% Intro %%%%%%%%%%%%%%%%%%%%%%%%%%%%%%%%%%%%%%

\section{Introduction}

% \textcolor{red}{Start with materials design to engineer custom functionality - need for theoretical tools to achieve this goal. The specific case of multiferroics and need for a single order parameter to determine the strength of coupling. Go onto multipoles.}. 

Rapidly evolving technologies create a continuous demand for solid-state materials with one or more functionalities tailored for specific applications. An important category of single-phase multifunctional materials is that of magnetoelectric multiferroics that possess spontaneous coexisting magnetic and ferroelectric orders \cite{Spaldin2005,Khomskii2006,Fiebig2016}. Although a number of materials are known to have these properties, those with potential for real-world applications belong to a smaller subset which exhibit substantial coupling between their magnetic and electric properties. Moreover, the multiferroics which exhibit strong interactions between electric and magnetic orders have very complex microscopic coupling mechanisms\cite{Ramakrishnan2019}, a deep understanding of which are essential to optimize their properties. \\ \par

In-depth understanding of multiferroicity on a microscopic level requires a combination of exhaustive experiments and theoretical analyses, often employing computational tools such as density functional theory (DFT). Neutron scattering is the method of choice to understand magnetic structure and quantum mechanical interactions at the fundamental level, and techniques using photons from the terahertz to X-ray regimes offer a wealth of complementary information. X-ray spectroscopic techniques (like X-ray magnetic circular dichroism, for example) are being regularly used to determine element specific electronic and magnetic properties in multiferroics. Ideally, one needs a technique which combines spectroscopy with diffraction to examine the long-range order of certain fine aspects of electronic and magnetic structure in a comprehensive manner.
\\  \par

%For example, it is believed that the mechanisms responsible for magnetoelectric coupling in the room temperature multiferroic BiFeO$_3$ produces a long-range magnetic spiral which can only be observed under very sensitive experiments\cite{Sosnowska1982}.

Resonant X-ray diffraction (RXD) is a technique that effectively combines diffraction with core-level absorption spectroscopy to observe long-range order in crystalline materials with element-specific electronic information. It has thus proved to be a useful tool to study fine details of magnetic arrangements and different types of magnetoelectric interactions in multiferroics over the years \cite{Wilkins2009,Walker2011,Windsor2015,Ramakrishnan2019}. In standard RXD experiments, one obtains the long-range ordered electronic properties, such as magnetic dipoles and anisotropies in the electron density distribution (also referred to as orbital order), within the material system. However, combining RXD with \textit{ab initio} calculations provides access to information regarding magnetic interactions, which are difficult to obtain using other experimental techniques \cite{Mannix2007, Lovesey2009, Dmitrienko2014, Ramakrishnan2017}. In particular, RXD is sensitive to long-range-ordered localized multipoles, including the exotic magnetoelectric multipoles \cite{Arima2005,Staub2009,Scagnoli2011}. Magnetoelectric multipoles are ground-state localized entities which simultaneously break parity and time-reversal symmetries \cite{Arima2005,Lovesey2009} and whose magnitudes and orientation in space can be calculated using DFT \cite{Spaldin2013}.  These multipoles have the appropriate symmetries to provide a single order-parameter in material systems lacking inversion and time-reversal\cite{Zimmermann2014}. More recently, it has also been suggested that they could be the order parameter for the pseudogap phase of high-temperature cuprate superconductors \cite{Shekhter2009,Scagnoli2011,Fechner2016}. Even though a lot of theoretical work has been performed on magnetoelectric multipoles \cite{Spaldin2013,Fechner2014,Thole2016,Meier2019}, the lack of model systems where their presence can be indisputably confirmed using RXD experiments has hindered the progress in this field. In this study, we take advantage of the recent progress in the FDMNES code \cite{Joly2001,Joly2012} which allows for a spherical tensor expansion of the scattering amplitudes contributing to a particular Bragg reflection as a function of energy, x-ray polarization and azimuthal angle, to disentangle the various multipolar contributions in h-YMO.
\\ \par

Hexagonal manganites with a general formula RMnO$_3$ (R = Y, In, Sc, Dy, Ho, Er, Tm, Yb, Lu) are one of the most studied classes of multiferroics. These are type-I multiferroics in which ferroelectricity sets in at a temperature $T_C$, which is well above the magnetic transition temperature $T_N$. Several types of antiferromagnetic orderings have been observed in compounds with different R atoms due to a complex interplay of geometrical frustration, spin-orbit coupling, lattice distortions and magnetic exchange interactions \cite{Fiebig2016,Brown2006}. One prominent system of this family, hexagonal YMnO$_3$ (h-YMO), crystallizes in the space group $P6_3/mmc$ at high temperatures and undergoes a geometrically driven ferroelectric transition around 1259 K, below which the symmetry changes to $P6_3cm$ \cite{Lilienblum2015}. The ferroelectricity in this material arises due to the buckling of the MnO$_5$ bi-pyramids \cite{Aken2004,Artyukhin2014}, which is dissimilar to the displacement of the B-site cations seen in orthorhombic ABO$_3$ perovskites. The onset of magnetic order takes place at $T_N \, \approx \, 71 K$, below which the Mn$^{3+}$ moments order in a non-collinear arrangement with magnetic symmetry \textit{P}6'$_3$\textit{cm}'.
\\ \par

Even though the ferroelectric and magnetic transitions in h-YMO occur independent of one another, anomalies are found in dielectric susceptibility at $T_N$ indicating strong magnetoelectric coupling \cite{Tomuta2001,Giraldo2021}. Magneto-elastic displacements of the Mn$^{3+}$ ions have also been observed below $T_N$ \cite{Lee2008}. The $d$-orbitals in the Mn$^{3+}$ ions are strongly anisotropic, and hence, canting of the magnetic moments perpendicular to the {\emph{ab}} plane are energetically favorable \cite{Solovyev2012}. An indication of this spin-canting along the \textbf{\emph{c}}-axis has been obtained from optical measurements \cite{Degenhardt2001}, but not in neutron scattering. \\ \par

In this article, we present our resonant X-ray magnetic diffraction studies to observe possible spin cantings and magnetoelectric multipoles in h-YMO. The article is organized as follows. In Sec. \ref{sec:exp}, the RXD experiment is described and basic results are analyzed. Detailed first-principles calculations of the RXD spectral profiles and magnetoelectric multipoles are described in Sec. \ref{sec:disc}, and the major outcomes are summarized in the concluding paragraphs.

%%%%%%%%%%%%%%%%%%% Experiment and Results %%%%%%%%%%%%%%%%

\section{Experimental Details}
\label{sec:exp}

%-------------------- Experimental ---------------------------

\subsection{Sample Preparation and Characterization}
\label{ssec:exp_samp}

Crystalline hexagonal YMnO$_3$ was prepared by the optical floating zone melting technique using a Cyberstar mirror furnace \cite{Lichtenberg2017}. Starting materials were powders of Y$_2$O$_3$ (Alfa Aesar, 99.99 \%, Lot B02X020) and Mn$_2$O$_3$ (MaTeck, 99.9 \%, Ch. 250708). The chemical composition of the Mn$_2$O$_3$ powder was checked by heating a small amount (about 70 mg)  up to 1100 \degree C at 10 \degree C/min under a flow of synthetic air in a thermogravimetric analyzer NETZSCH TG 209 F1 Libra or NETZSCH STA 449 C Jupiter. The dwell time at 1100 \degree C was 5 min followed by a cooling down to 100 \degree C at -30 \degree C/min. A relatively large and step-like weight loss was observed in the temperature range of about 900 - 1000 \degree C, followed by a constant weight at higher temperatures. The chemical composition of Mn$_2$O$_3$ was confirmed since the observed weight loss above about 900 \degree C corresponds precisely to the mass loss which is expected from the well-known transformation of Mn$_2$O$_3$ into Mn$_3$O$_4$. Further details, results as well as pictures are presented in the thermogravimetry section of Ref. \onlinecite{Lichtenberg2017}.

\subsection{X-ray Diffraction}
\label{ssec:exp_xrd}

The experiments were carried out at the endstation RESOXS \cite{Staub2010} at the X11MA beamline \cite{Flechsig2010} of the Swiss Light Source. The single crystal was mounted such that the [001] direction was along the horizontal scattering plane. Linear horizontal ($\pi$) and vertical ($\sigma$) polarized X-rays were focused at the sample with a spot-size of 130 $\times$ 50 $\mu m$. The beamline produced monochromatic x-rays with an energy resolution of about 0.15 eV and at the Mn $L_{2,3}$ edges. The sample was manually rotated in-situ with an accuracy of $\pm 3 \degree$ for the azimuthal angle ($\Psi$) scans. Reciprocal space scans along $(0,0,L)$ were performed as a function of energy and temperature, with $\pi$ polarized x-rays. \\ \par

%%%%%%%%%%%%%%%%%%%%%%% Results %%%%%%%%%%%%%%%%%%%%%%%%%%%%%%%

\section{Results and Interpretation}
\label{sec:res}

\begin{figure*}[ht]
\includegraphics[width=1.0\textwidth]{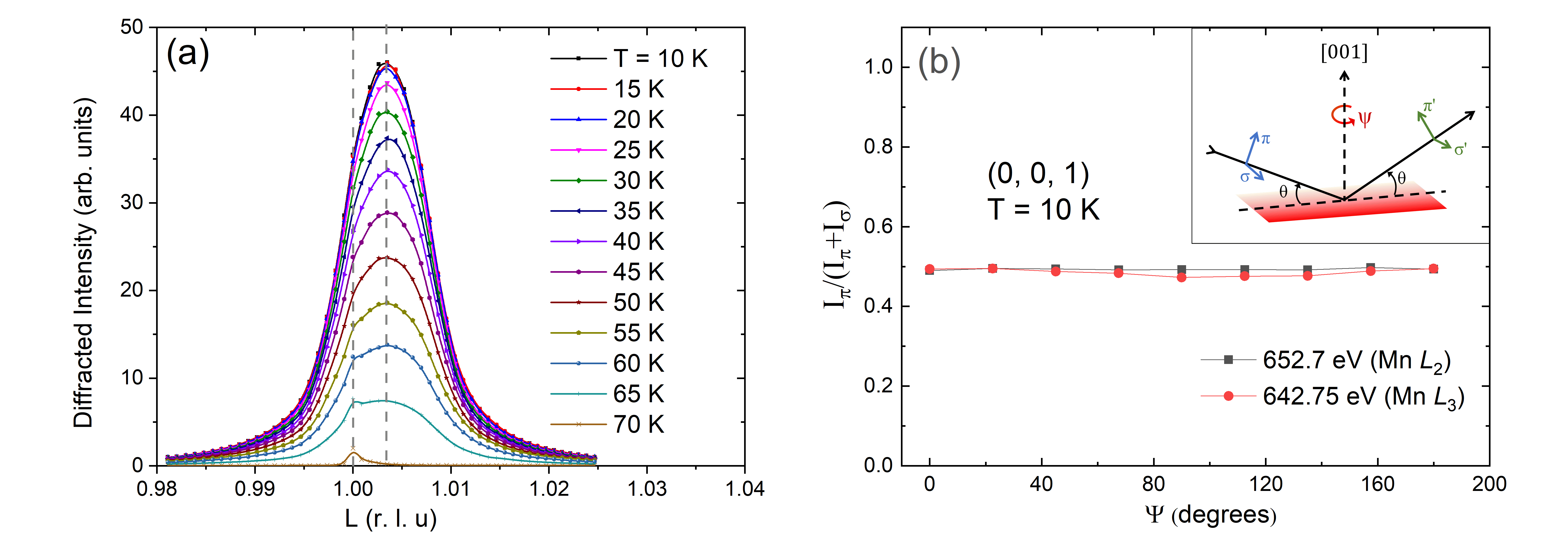}
\caption{(a) Figure showing the magnetic $(0,0,1)$ diffraction peak in h-YMO for various temperatures, measured at E = 642.75 eV (Mn $L_3$ edge) using $\pi$-polarized x-rays. The dashed vertical lines indicate the positions of the magnetic peak which undergoes refraction at the Mn L$_3$ edge and is consequently not centered exactly at $L=1$, and the residual peak at 70 K which originates from the $\lambda/2$ leakage of the monochromator diffracting off the symmetry allowed $(0,0,2)$ Bragg reflection. (b) Azimuthal dependence of the ratio $I_\pi/(I_\sigma+I_\pi)$ in h-YMO at $T$ = 10 K, for the $L_3$ and $L_2$ edges of Mn. The error bars are within the symbols, and the solid lines are a guide to the eye. The inset shows the diffraction geometry.}
\label{fig:tdep1}
\end{figure*}

In h-YMO, the $(0,0,1)$ Bragg reflection is forbidden according to the $P6_3cm$ space-group, but a strongly resonant diffraction signal was observed below the N\'eel temperature  $T_N$.  Fig. \ref{fig:tdep1} shows the reciprocal space scans along $(0,0,L)$ showing an intense diffraction peak. The intensity of the reflection was found to reduce with increasing temperature and eventually go to zero at $T_N$, thereby proving its magnetic origin.  \\ \par

%\textcolor{red}{Reduce the number of equations, use the equation for the ratio in case of absence of azimuthal dependence. Absence of threefold braking since this signal hasn't been observed for neutron scattering. Show that this comes from the AFM canting.}
\subsection{Origin of the forbidden Bragg reflection}

To understand the nature of the magnetic form factor contributing to this Bragg reflection, the dependence of the scattering intensity on x-ray polarization and azimuthal angle was investigated. Fig. \ref{fig:tdep1} shows the ratio $I_\pi/(I_\sigma+I_\pi)$ of the $(0,0,1)$ reflection as a function of the azimuthal angle $\Psi$. The intensity is independent of the azimuthal angle within experimental accuracy, and has equal intensities in both polarization channels. The reflection also shows identical spectral shapes across the Mn $L_{2,3}$ edges with both $\sigma$- and $\pi$-polarized x-rays. The precise nature of the magnetic moments (and/or other scattering tensors) contributing to this reflection can be understood by evaluating the structure factor. The structure factors, for a resonant Bragg reflection, can be written in the most general form as -

\begin{equation}
\label{eq:strfac}
S(h,k,l) = \sum_n{f_n (E) e^{i 2\pi (h\hat{x}+k\hat{y}+l\hat{z})\mathbf{\cdot r}_n}}
\end{equation}

where $h, k, l$ are the Miller indices and $n$ runs over all the resonant atoms (Mn) in the unit cell. In the magnetic unit cell (which is same as the structural unit cell in h-YMO), there are six Mn atoms - three atoms $(n=1,2,3)$ at $z=0$ and three $(n=4,5,6)$ at $z=0.5$. For the $(0,0,1)$ Bragg reflection with $h=k=0$, Eq. (\ref{eq:strfac}) reduces to - 

%The energy dependence of $f_n$ is ignored in Eq. (\ref{eq:strfac}).

%\begin{equation}
%\label{eq:strfac2}
%S(0,0,1) = \sum_n{f_n  e^{i 2\pi (z_n)}}
%\end{equation}

% where $z_n$ is the $z$-coordinate of the $n^{th}$ Mn atom. The structure factor thus assumes the form  

\begin{equation}
\label{eq:strfac3}
S(0,0,1) = (f_1 + f_2 + f_3) - (f_4 + f_5 + f_6)
\end{equation}

Since the scattering intensity is equal for both $\sigma$ and $\pi$ polarizations, the same structure factor is valid for both polarizations. As a first approximation, we look at the scattering terms within the form factor $f_n$, originating from the $E1E1$ process alone \cite{Hill1996}. Here $E1$ refers to an electric dipole transition between the core and valence atomic levels involved ($E$ stands for electric, and the number denotes the $\Delta l$ between the atomic states. $E2$ and $M1$ would thus refer to an electric quadrupole and a magnetic dipole transition, respectively. Since RXD is a two-photon process, we always look at combination of two transitions.) Equal intensities in both polarization channels ($I_\sigma=I_\pi$) and the absence of azimuthal dependence implies that 

\begin{equation}
\label{eq:fapprox}
f_n \propto  m_n^{z} sin\theta
\end{equation}

where $m_n^{z}$ is the spin-component along the $\mathbf{\hat{z}}$ direction, also indicating scattering only in the rotated polarization channels ($\sigma \rightarrow \pi'$ and $\pi \rightarrow \sigma'$)\cite{Joly2009,Ramakrishnan2017}. Due to the negative sign in Eq. \ref{eq:strfac3}, only the antiferromagnetic (AFM) component of the spins contributes to the structure factor. In other words, the $(0,0,1)$ reflection directly measures the AFM spin canting along the \textbf{c}-axis of the crystal. An indication for such a spin-canting has been reported earlier from optical second-harmonic generation (SHG) studies \cite{Degenhardt2001}. Polarized neutrons are not sensitive to spin components along the Bragg wavevector. Hence, this AFM-canting of the Mn moments along the \textbf{c}-axis does not contribute to any $(0,0,L)$ type of reflections in neutron diffraction experiments. It should be noted that the magnetic $(0,0,1)$ reflection which has been observed in neutron diffraction of h-HoMnO$_3$ \cite{Lonkai2004} originates from the long-range ordering of the Ho$^{3+}$ magnetic moments. This is unlike the case of Y$^{3+}$ ions, which do not have ordered magnetic moments\cite{Lonkai2002}.

%---------- Spectral Shape -----------------

\subsection{Spectral shape evolution with temperature}
\label{ssec:rxd_spec}

\begin{figure*}[ht]
\includegraphics[width=1.0\textwidth]{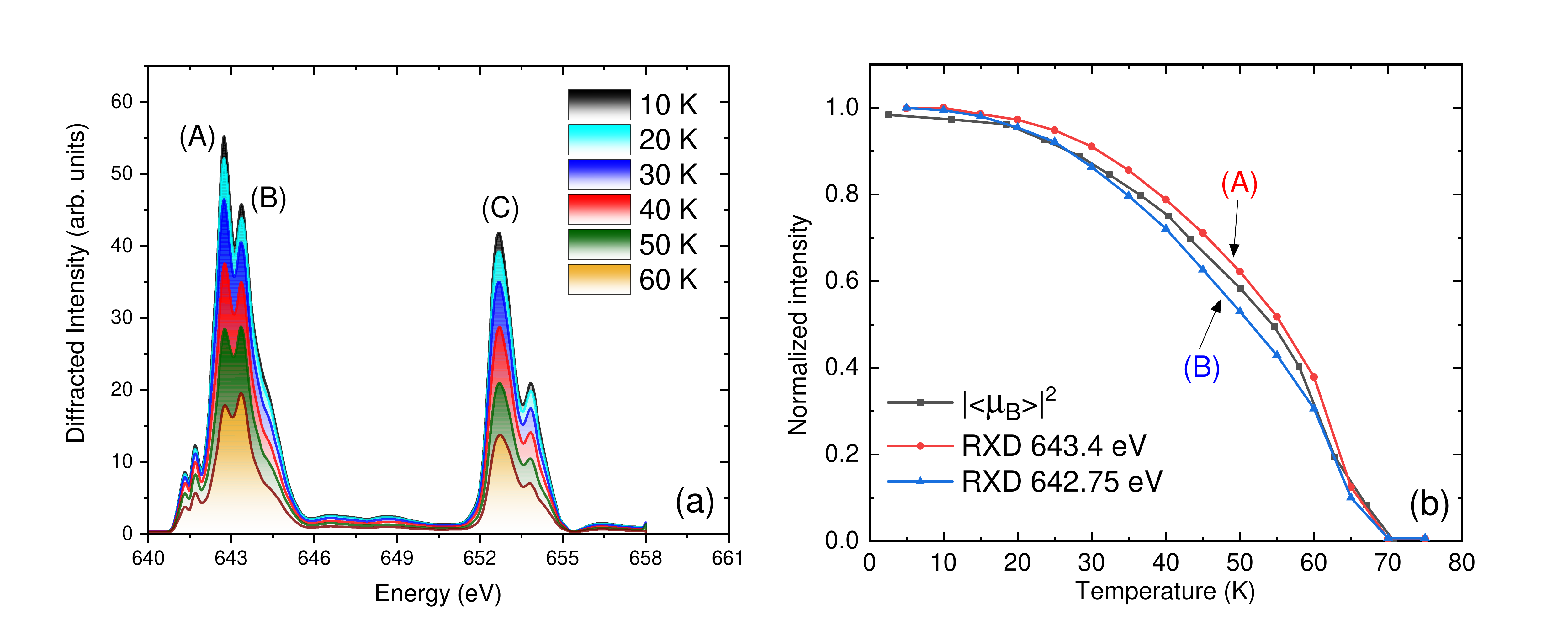}
\caption{(a) The spectral intensity profile of the (0,0,1) reflection at the Mn $L_{2,3}$ edges at various temperatures. (b) The temperature dependence of the diffraction intensity at the energies indicated A and B, and the square of the magnetic moment obtained in neutron diffraction (adapted with permission from Ref. \onlinecite{Munoz2000}. Copyrighted by the American Physical Society). The intensities have been normalized to unity at the lowest temperature, and the temperature axis has been adjusted so that the $T_N$ of both experiments coincide. The solid lines are guides to the eye.}
\label{fig:efixq}
\end{figure*}

In the expression for the structure factor for the $(0,0,1)$ magnetic Bragg reflection presented earlier, the energy dependence of the form factor was ignored. However, near an atomic absorption edge, the form factor $f_n(E)$ has a strong dependence on energy. The resulting energy dependence of the diffraction intensity at the absorption edge, called the spectral profile, contains detailed information regarding the symmetry and magnetoelectric interactions \cite{Ramakrishnan2017}. Hence, we measured the spectral profile of the magnetic $(0,0,1)$ reflection for several temperatures (see Fig. \ref{fig:efixq}). The spectra were obtained by integrating the intensity of the reciprocal space scans of the $(0,0,1)$ reflection at every energy point around the Mn $L_{2,3}$ edges. \\ \par

A striking observation is that the shape of the spectrum changes with temperature. For example, the relative intensity of the two peak-like features A and B at the Mn $L_3$ edge vary with temperature. This is also clearly seen in measurements of the scattered intensity for different energies with finer temperature steps. Fig. \ref{fig:efixq} shows the intensity at each temperature normalized with respect to the intensity at the base temperature, for two different energies (corresponding to A and B Fig. \ref{fig:efixq}(a)). A detailed discussion about the origin of this observation is provided in the following section. \\ \par

%%%%%%%%%%%%%%%%% Discussion %%%%%%%%%%%%%%%%%%%%

\section{Discussion and Calculations}
\label{sec:disc}

%----------------- Magnetic Symmetry ------------------------
\subsection{Antiferromagnetic canting}
\label{ssec:disc_mag}

The series of hexagonal manganites feature a wide variety of magnetic configurations whose origins have been extensively studied by a variety of techniques. Resonant x-ray diffraction at the Mn $L_{2,3}$ edges is sensitive to electronic ordering phenomena local to the Mn$^{3+}$ ions. The $(0,0,1)$ Bragg reflection, which is forbidden according to the $P6_3cm$ space group, shows strong resonant scattering below $T_N$. As seen in the previous section, this Bragg peak originates from the antiferromagnetic canting of spins along the \textbf{c}-axis. This is yet another demonstration of the ability of RXD to investigate features like spin canting with high sensitivity. However, it must be mentioned that it is not possible to obtain a quantitative estimate of the canting angle in this particular case. This type of spin canting is allowed under the magnetic space group $P6'_3cm'$. \\ \par

Even though there have even been indications of such a spin-canting from optical SHG experiments\cite{Degenhardt2001}, it had not yet been reported in any scattering experiments. Spin-cantings in ME systems are usually described by the antisymmetric exchange mechanism based on relativistic Dzyaloshinkii-Moriya (DM) interactions. It would be of fundamental interest to understand whether we can control the strength of the DM interactions and thereby, the canted moments using strain. Hence, one needs to repeat these experiments on differently strained epitaxial films, where the magnetic ordering temperature has been reported to change as a function of strain \cite{Wu2013}. \\ \par

%%----------------Spectral Shapes-----------------

\subsection{Changes in Spectral Shape}
\label{ssec:disc_spec}

In \ref{ssec:rxd_spec}, it was seen that the spectral shape of the $(0,0,1)$ Bragg reflection was different for different temperatures. Changes in symmetry at the site of the resonant atom or its position in the unit cell can, in principle, alter the local electronic distribution affecting the RXD spectra \cite{Ramakrishnan2017}. However, it has been observed experimentally that movement of the Mn atoms within the unit cell occurs only at temperatures close to $T_N$, and, no structural changes have been observed below 40 K \cite{Lee2008}. Hence, at temperatures below 40 K, any kind of atomic motion induced changes in spectral shape can be ruled out. Fine changes in the magnetic structure like the canting angle can also be ruled out since the temperature dependence of the (0,0,1) reflection coincides with the total magnetic moment in the system as observed with neutrons (see Fig. \ref{fig:efixq}). \\ \par 

\begin{figure}[h]
\includegraphics[width=0.45\textwidth]{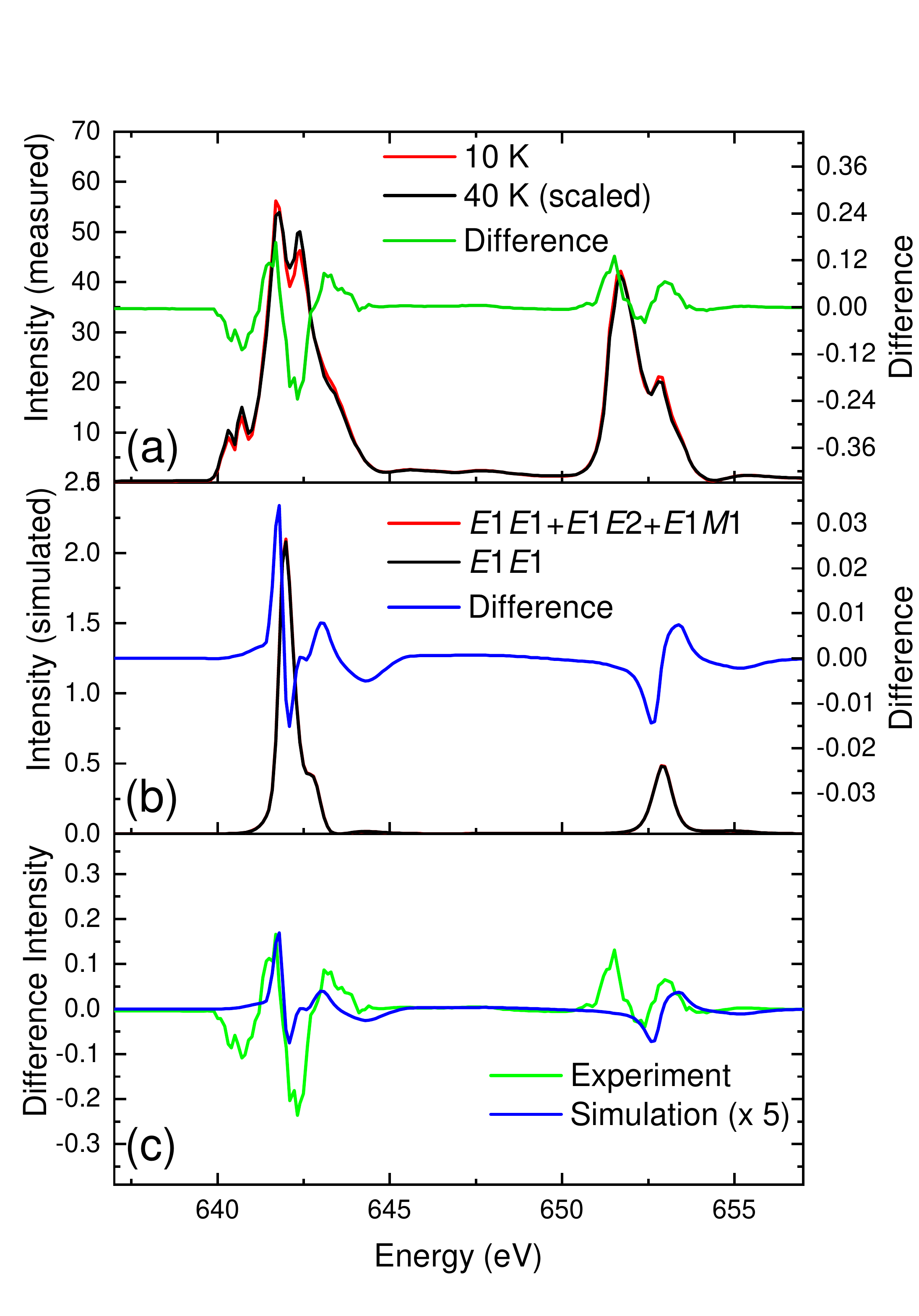}
\caption{(a) The normalized spectral shapes of the magnetic $(0,0,1)$ reflection obtained at T = 10 K and T = 40 K.The intensities were normalized such that the difference spectrum averages to zero when integrated over the given energy range. (b) Spectral shapes obtained with and without contributions from ME multipoles, calculated using FDMNES. In addition to $E1E1$, the $E1E2$ and $E1M1$ transition processes were also included in the calculation for the ME multipoles. The difference between the two profiles is also shown in both panels. (c) The comparison of the experimental and calculated difference spectral profiles.}
\label{fig:specDiff}
\end{figure}

To understand other possible causes for this spectral change, we need to revisit the approximation made to arrive at Eq. \ref{eq:fapprox}, where we limited the form factor $f_n(E)$ to scattering terms originating from the $E1E1$ process. Higher order electric ($E1E2$, $E2E2$) and mixed electric-magnetic processes ($E1M1$) have been found to contribute to resonant X-ray scattering in several correlated electron materials ($E2$: electric quadrupole transition, $M1$: magnetic dipole transition) \cite{Matteo2005}. Since scattering terms originating from different resonant processes can have different amplitudes and phases as a function of energy, the final spectral shape is the result of interference of all such contributions-

\begin{equation}
\label{eq:ftotal}
\begin{split}
f_n (E) &  \propto f_n^{E1E1} (E) + f_n^{E1E2} (E) \\
        & + f_n^{E1M1} (E)+ f_n^{E2E2} (E)
\end{split}
\end{equation}

where $f_n^{E1E1}(E) \propto  m_n^{z} sin\theta$ (given by Eq. \ref{eq:fapprox}). The higher order terms $f_n^{E1E2} (E)$, $f_n^{E1E2} (E)$, and $f_n^{E1E2} (E)$ denote the combined form factor of all allowed multipoles from the respective processes. Only those multipoles which are long-range ordered with the Fourier component along the $(0,0,1)$ wavevector contribute to this reflection \cite{Lovesey2009,Scagnoli2011,Staub2010}. Moreover, since there is no dependence of the scattering intensity on the azimuthal angle, the relevant atomic multipoles should also be symmetric with respect to any rotation about the \textbf{c}-axis. \\ \par

To investigate interference of one or more atomic multipoles in our experimental spectra, we look at the difference of the normalized spectra at 10 K and at 40 K. Fig. \ref{fig:specDiff} shows the the normalized spectra of the $(0,0,1)$ reflection measured at 10K and 40K, and the difference spectral profile. The difference spectrum is obtained after normalizing the two spectra so that they have equal weight when integrating the intensity over the edges. The difference resembles a scenario when there is interference between different scattering terms \cite{Dmitrienko2014,Sessoli2015}. A quantitative evaluation is not feasible with the computational tools currently available. However, we employ a combination of DFT and phenomenology to provide a semi-quantitative description of this interference using a model of resonant scattering from the magnetoelectric multipoles. \\ \par

%---------------- Ab initio Calculations -------------------
\subsection{\textit{Ab initio} Calculations}
\label{ssec:abinitio}

The presence of higher order multipoles in the scattering signal can be addressed using the \textit{ab initio} FDMNES code \cite{Joly2001,Joly2009}. The package uses a given crystal and starting magnetic structure to compute the spin-polarized density of states of a given material using density functional theory (DFT). Following this, the calculations for spectra for x-ray absorption and x-ray diffraction are performed. The code enables one to choose the transition processes for which the absorption or diffraction spectra are calculated. FDMNES does not compute the relaxed crystal structure, and hence, we used the crystal structure provided in Ref. \onlinecite{Munoz2000}. The magnetic structure corresponding to the space group $P6'_3cm'$ given in Ref. \onlinecite{Brown2006} was used. To correlate with our findings, an antiferromagnetic canting of 1$\degree$ was added to this input magnetic structure. The fully relativistic calculations were performed in the self-consistent mode using multiple scattering approach to obtain the magnetic ground state\cite{Joly2001,Joly2012}. A cluster radius of 4\AA \, and a uniform broadening of 0.1 eV were used for all calculations. The polarization resolved scattering intensities for the $(0,0,1)$ Bragg reflection were calculated for $E1E1$ and a combination of $E1E1$, $E1E2$ and $E1M1$ processes, including correction factors for self absorption. Further details and a sample code are given in Ref. \onlinecite{RamakrishnanThesis2017}. Due to inherent limitations in the estimation of core-hole effects and other interactions in the excited-state, the multiplet structure of the partially filled $3d$ orbitals are not accurately computed. Hence, the shapes of the spectra at the $L_{2,3}$ edges of Mn are not reproduced. However, we can get a semi-quantitative estimate of the relative scattering contributions from the magnetic canting and other higher order multipoles. \\ \par

For simplicity, we limit our calculations to the $E1E1$, $E1E2$ and $E1M1$ processes. Even though the latter processes are much weaker than $E1E1$, they can interfere amongst themselves giving visible changes in the spectral profiles. Including this interference, the total scattering intensity for a combination of the above processes can be approximated as -

\begin{equation}
\label{eq:Itot}
\begin{split}
I^{tot} (E) & \propto |f_n^{E1E1} (E) + f_n^{E1E2} (E) \\ 
& + f_n^{E1M1} (E)+ f_n^{E2E2} (E)|^2 \\ \\
& \approx |f_n^{E1E1} (E)|^2  +  |f_n^{E1E1} (E)f_n^{E1E2} (E)| \\
& +  |f_n^{E1E1} (E)f_n^{E1M1} (E)|
\end{split}
\end{equation}

%%% Start revision from here %%%

The squares and combinations of higher order scattering terms can be neglected since they are generally too weak to be detected in such an experiment. The interference terms $|f_n^{E1E1}(E) .f_n^{E1E2}(E)|$ and $ |f_n^{E1E1}(E).f_n^{E1M1} (E)|$ enable us to observe the weak scattering from the higher order multipoles. In the calculation using FDMNES, we can selectively calculate the scattering intensities from each process or any combinations of these (to account for any interference) \cite{Joly2009}. Thus, we calculate the scattering intensities in case of (i) a $E1E1$ transition process alone, and (ii) combination of $E1E1$, $E1E2$ and $E1M1$ transition processes. We focus only on the scattered intensity in the rotated light channels for the $(0,0,1)$ reflection, in accordance with experimental observations. The intensity in these channels for case (i) is exclusively the scattering due to the AFM canting of the magnetic dipole moments along the hexagonal \textbf{c}-axis. For case (ii), the diffraction amplitudes from the higher-order multipoles interfere with the strong scattering signal from the canted AFM dipoles. On subtracting the spectra obtained for cases (i) and (ii) above, we observe a clear difference, of the order of a few percentage of the total diffraction intensity. As we can see, the calculations do not reproduce all features of the experimental spectra given in Fig. \ref{fig:specDiff}. The spectrum is shifted on the energy axis due to the inaccurate determination of the Fermi energy of the system in the presence of a core-hole. This difference spectrum along with the intensity profiles for calculations (i) and (ii) described above are plotted in Fig. \ref{fig:specDiff}. The difference spectrum obtained experimentally can now be compared with the calculated one [see Fig. \ref{fig:specDiff}(c)]. The mismatch between experiment and calculation is likely due to the fact that the effects like localization of electronic states in presence of the core-hole is not well accounted for in DFT-based calculations of $3d$ systems. \\ \par

%---------- Magnetoelectric multipoles ---------------

\subsection{Multipolar Analysis}
\label{ssec:calc_mult}

\begin{figure}[h]
\includegraphics[width=0.45\textwidth]{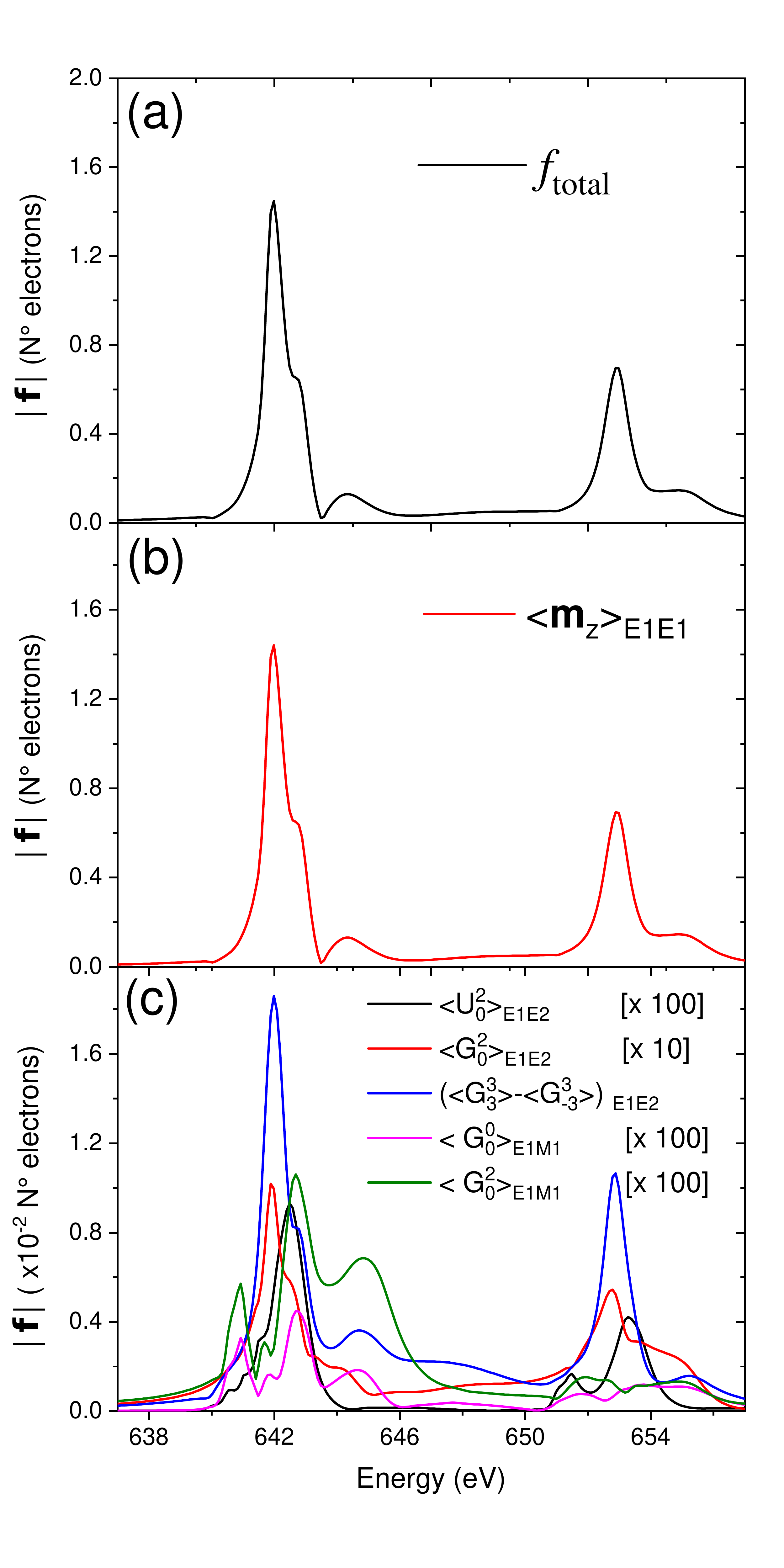}
\caption{Calculated spectral intensities for (a) the overall form factor of the $(0,0,1)$ reflection in h-YMO obtained following interference of magnetic dipole and magnetoelectric multipoles, (b) the magnetic dipole term, and (c) the magnetoelectric $\langle G^K_Q \rangle$ multipoles at the Mn $L_{2,3}$ edges.}
\label{fig:me}
\end{figure}

In Sec. \ref{ssec:abinitio}, we established that the anomalous evolution of the spectral shapes with temperature can be explained by considering interference of scattering signals from the canted AFM dipoles and higher-order multipoles. The FDMNES code also allows the expansion of the scattering tensor in cartesian and/or spherical tensors, and to obtain the contribution of individual atomic multipoles. We expand the intensities as spherical tensors, and use the notation introduced in Refs. \onlinecite{Lovesey2005,Lovesey2009}. The atomic tensors derived from spherical harmonics are denoted by $\langle X^K_Q \rangle$, where $X$ is the tensor type ($T$: parity-even and non-magnetic; $U$: parity-odd and non-magnetic, called polar multipoles; $G$: parity-odd and magnetic, called magnetoelectric multipoles), $K$ is the rank of the tensor (0: monopolar, 1: dipolar, 2: quadrupolar etc.) and $Q$ is the projection of the tensor on the chosen basis. The multipolar contributions to the scattering intensity are shown in Fig. \ref{fig:me}. The figure shows the square of the form factor for all the nonzero multipoles obtained from $E1E1$, $E1E2$ and $E1M1$ processes as a function of energy, in the FDMNES calculation for h-YMO. The strongest scattering term is the \emph{magnetoelectric octupole}, which is represented as $ \langle G^3_3 \rangle -\langle G^3_{-3} \rangle$. This spherical octupole resembles an \textit{f} orbital which is symmetric with respect to rotations about the principal axis. In our calculations, the magnitude of the form factor corresponding to this octupole is about $1\%$ of scattering from the magnetic dipole. One should note that even though the scattering intensities from the individual multipoles shown in the figure are small, they interfere with each other affecting the overall spectral shape significantly. Certain multipoles contribute to scattering in both the $E1E2$ and the $E1M1$ processes, albeit with different spectral shapes. For the Mn $L_{2,3}$ edges in h-YMO, the intensities resulting from the $E1E2$ process are generally stronger at these energies compared to those from $E1M1$, exemplified by the intensity of the magnetoelectric quadrupole $\langle G^2_0 \rangle$. However, to our knowledge, no experimental evaluation of the overall cross-sections of these two processes in RXD has been done to date. Yet another quantity of tremendous interest is $\langle G^0_0 \rangle$, which is a magnetic rank zero tensor. This entity, referred to as the magnetoelectric monopole or the magnetic charge, is fundamentally different from the monopole forbidden by classical electromagnetism  \cite{Spaldin2013}. It has nonzero intensity in our calculations, even though its contribution is weaker compared to the other multipoles. \\ \par

%\textcolor{red}{Discuss implications on the symmetry understanding in hexagonal manganites.}
%In order to get a basic understanding of how the spectrum changes with increasing temperature, we calculated the total intensity of all the scattering terms as a function of the spin moment on the Mn atoms. Fig. \ref{fig:spin_calc} shows the RXD spectra as a function of the initial spin configuration of Mn used as the input for the DFT calculation. It was observed that the calculations become less reliable as the spin configuration on the atoms is forced to a value different to the actual ground state moment. Hence, such an estimation of the temperature dependence of the RXD spectra is not possible. 

%\begin{figure}[h]
%\includegraphics[width=0.4\textwidth]{images/Spin_calc.png}
%\caption{Calculated intensities for the case of artificially modifying the spin moment on the Mn atoms. For initial electronic configurations other than the nominal ground state, the calculations do not give reliable results.}
%\label{fig:spin_calc}
%\end{figure}

\begin{figure*}[ht]
\includegraphics[width=0.7\textwidth]{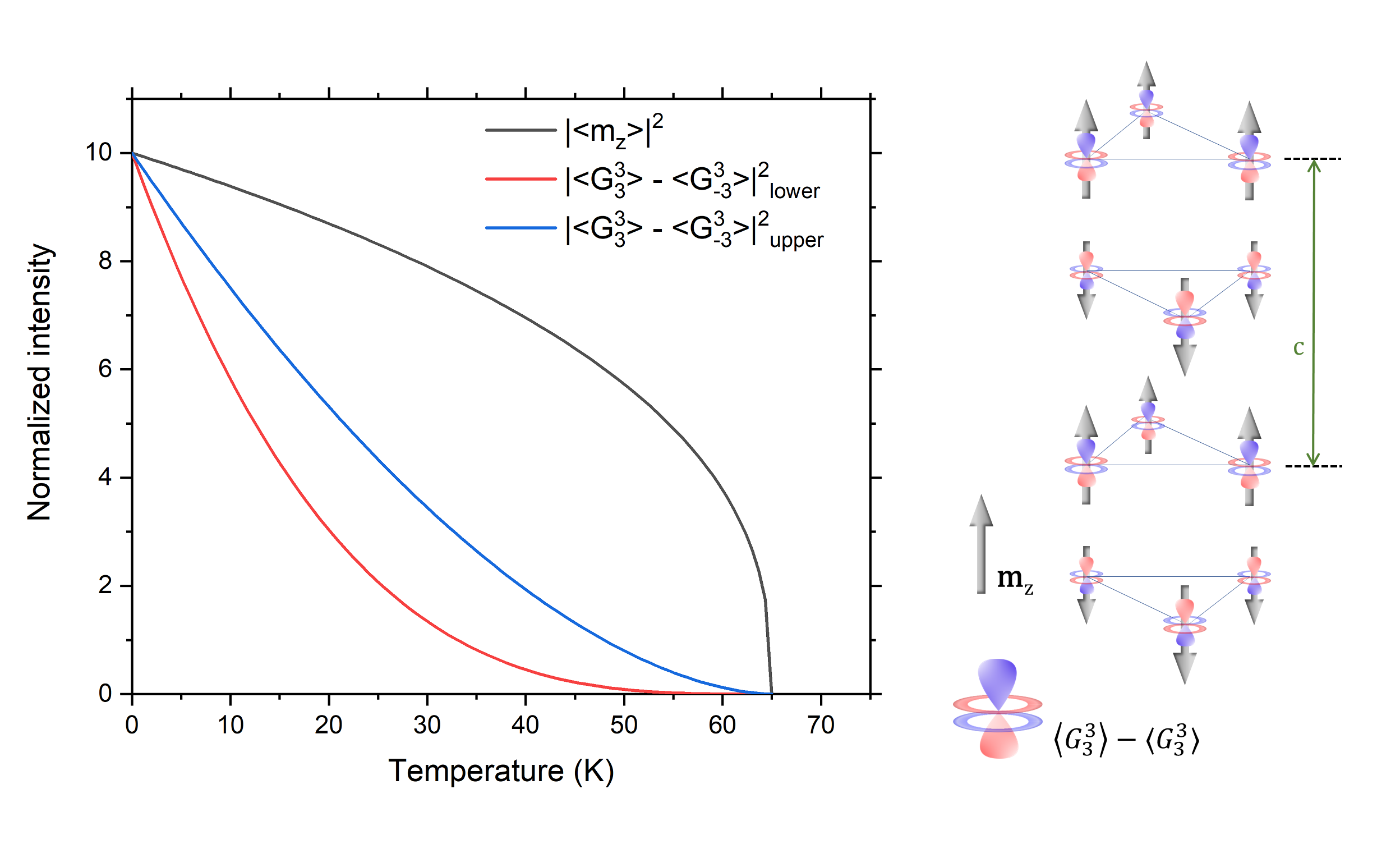}
\caption{The comparison of temperature dependence of the magnetic scattering upper and lower bounds of scattering intensity from the magnetoelectric octupole $\langle G^3_3\rangle - \langle G^3_{-3} \rangle$, according to our model. The arrangement of the canted part of the magnetic moment and the magnetoelectric octupoles is also shown.}
\label{fig:tdepME}
\end{figure*}

One key aspect that has not been discussed so far concerns how interference of magnetoelectric terms leads to a different spectral shape at 40 K compared to 10 K. Ideally, one should calculate the RXD spectra as a function of temperature. There is a dearth of computational tools to quantitatively simulate the spectra for temperatures other than absolute zero. DFT-based methods are usually employed to deal with the ground state of a system. Due to the above reasons, we can only provide a phenomenological explanation for the observed temperature dependence of the RXD spectra. The temperature dependence of a purely magnetic term contributing to scattering can be measured in an experiment. Since the intensity of the $(0,0,1)$ Bragg reflection in h-YMO is heavily dominated by magnetic scattering, we can assume its temperature dependence to follow that of a pure magnetic dipole. Upon fitting the normalized intensity with a mean-field model $I \propto (T_N - T)^{2\beta_{mag}}$, we obtain $\beta_{mag} \approx 0.38$, where $\beta_{mag}$ is the critical exponent of the magnetic scattering from the canted AFM moments. Magnetoelectric multipoles, on the other hand, are products of spatial and spin-density terms \cite{Matteo2005,Spaldin2013}. Therefore, they have distinct temperature dependences, based on their actual tensorial form. Since the polar toroidal octupole is by far the strongest higher-order scattering term, we ignore the other multipoles to simplify our analysis. This octupole, which is a product of the spin density term and a spatial term to the power of two \cite{Matteo2005} can be approximated as:

\begin{equation}
\vert \langle G^3_3 \rangle - \langle G^3_{-3} \rangle \vert (T)  \, \propto \,  \mu(T) \, r^2(T)   
\label{eq:tdep_oct}
\end{equation}

where $\mu(T)$ and $r(T)$ are the temperature-dependent spin density and spatial terms. From literature, for spatially dependent electric polarization (like in ferroelectric materials) which depends linearly on $r(T)$, the value of the critical exponent $\beta_r$ (critical exponent for the spatial dependence) falls within the range of 0.24 to 0.62 \cite{Kadanoff1967,Say2010,Sarikaya2013}. From Eq. \ref{eq:tdep_oct}, $\beta_{oct}$ = $\beta_{mag} + 2 \beta_r$ and hence, we can approximate the value of the overall critical exponent $\beta_{oct}$ for the octupoles to be between 0.86 and 1.62. Based on these critical exponents, we can model the scattering intensity as a function of temperature for the magnetic dipole moments and octupoles as shown in Fig. \ref{fig:tdepME}. It is clear that the scattering contribution from magnetoelectric octupoles decreases at a comparatively higher rate with increasing temperature. Hence, the overall spectral shape is expected to change as a function of temperature and, for $T\approx40$ K, one can assume that there is a relatively smaller contribution from $f_n^{E1E2} (E)$ and $f_n^{E1M1} (E)$ compared to $f_n^{E1E1}(E)$.
\\ \par

In RXD, the dipole-quadrupole $E1E2$ process is usually invoked in studies involving the pre-edge region of $K$ edges of transition metals (where the $E2$ excitation $1s \rightarrow nd$ probes the partially filled $d$-states), or the $L$ edges of rare-earths (where the $E2$ excitation $2p \rightarrow nf$ probes the $f$-states). The fact that we find a measurable cross-section for the $E1E2$ process at the Mn $L_{2,3}$ edges is very interesting from a fundamental perspective. For example, this could be due to a strong $d-f$ hybridization leading to an $f$-like character of the final states.  Note that this effect is visible due to the small spin canting leading to an effective \textbf{c} axis projection of the dipole moment that is approximately two orders of magnitude reduced in strength.\\ \par

\begin{comment}
Whether this is, in fact, due to the inter-atomic hybridization between the Mn $d$-orbitals and the low lying empty $f$-orbitals of Y remains to be investigated.
\end{comment}

Earlier reports of changes in spectral shapes in resonant diffraction have occurred in systems with either atomic motion or macroscopic changes like spin rotation \cite{Staub2017}. In the absence of any of these observable changes, an observable change in the spectral shapes due to magnetoelectric multipoles is the most probable explanation. More investigations are needed to understand this phenomenon, complemented by dedicated theoretical and computational studies, ultimately to the comprehensive understanding of diffraction anomalous fine structure (DAFS) over large energy ranges in correlated electron materials.  \\ \par

%%%%%%%%%%%%%%%%%%%%%%%%%%% Summary %%%%%%%%%%%%%%%%%%%%%
\section{Summary}

We investigate the $(0,0,1)$ Bragg reflection below $T_N$ in a single crystal of hexagonal YMnO$_3$ using resonant X-ray diffraction (RXD). Following a detailed examination of the dependence of diffraction intensity on X-ray polarization and azimuthal angle, we can conclude that this reflection, which is forbidden according to the $P6_3cm$ space-group, originates from an antiferromagnetic canting of the Mn$^{3+}$ magnetic moments perpendicular to the crystallographic \emph{ab} plane. We also observe that the shape of RXD spectra changes for different temperatures. Using \textit{ab initio} calculations and phenomenological arguments, we discuss this behavior from the perspective of the interference between scattering from the magnetic dipole and parity-odd atomic multipoles on Mn ions. A detailed microscopic theory on the behavior of magnetoelectric multipoles at temperatures above absolute zero is necessary to validate our hypothesis and, in general, to expand the scope of this method in the broader field of multiferroics.

%%%%%%%%%%%%%%%%%%%%%%%%%%% Acknowledgements %%%%%%%%%%%%%%%%%%%%%
\section*{Acknowledgements}
The authors are grateful to N. A. Spaldin and M. Fiebig for insightful discussions and comments on the manuscript. We thank J. -G. Park and Seongsu Lee for providing the structural data published in Ref. \onlinecite{Lee2008}, and A. Mu\~noz for permission to reuse data from Ref. \onlinecite{Munoz2000}. The RXD experiments were carried out at X11MA beamline of the Swiss Light Source, Paul Scherrer Institut, Villigen, Switzerland. The authors thank the X11MA beamline staff for experimental support. The financial support of the Swiss National Science Foundation (SNSF) is gratefully acknowledged (Projects No. CRSII2\_147606 and No. 200020\_159220). E.M.B. and U.S. acknowledge financial support from NCCR MUST (No. 51NF40-183615) and NCCR MARVEL (No. 182892), a research instrument of the SNSF, and funding from the European Community’s Seventh Framework Program (FP7/2007-2013) under Grant No. 290605 (COFUND:PSI-FELLOW). F.L. thanks Barbara Scherrer, former member of the division Nonmetallic Inorganic Materials of the Department of Materials of the ETH Zurich, for her assistance concerning thermogravimetry with the system NETZSCH STA 449 C Jupiter.

%%%%%%%%%%%%%%%%%%%%%%%%%%%%%%%%%% Refs (BiBTeX) %%%%%%%%%%%%%%%%%%%%%%%%%%%%%%%
\bibliographystyle{unsrt}
\setcitestyle{numbers}
\bibliography{ymo}

\end{document}